\documentclass[prb,amsmath,amssymb,superscriptaddress,twocolumn]{revtex4}
\usepackage{braket}
\usepackage{graphicx}
\usepackage{bbm,color,ulem}
\DeclareMathAlphabet{\mathpzc}{OT1}{pzc}{m}{it}
\begin{document}
\title{Plasmon-pole approximation for many-body effects in extrinsic graphene}
\author{E.\ H.\ Hwang}
\affiliation{Condensed Matter Theory Center and Joint Quantum Institute, Department of Physics, University of Maryland, College Park, Maryland 20742-4111 USA}
\affiliation{SKKU Advanced Institute of Nanotechnology and Department of Nano Engineering, Sungkyunkwan University, Suwon, 16419, Korea}
\author{Robert E.\ Throckmorton}
\affiliation{Condensed Matter Theory Center and Joint Quantum Institute, Department of Physics, University of Maryland, College Park, Maryland 20742-4111 USA}
\author{S.\ Das Sarma}
\affiliation{Condensed Matter Theory Center and Joint Quantum Institute, Department of Physics, University of Maryland, College Park, Maryland 20742-4111 USA}
\date{\today}
\begin{abstract}
We develop the plasmon-pole approximation (PPA) theory for calculating the carrier self-energy of extrinsic
graphene as a function of doping density within analytical approximations to the $GW$ random phase approximation
($GW$-RPA).  Our calculated self-energy shows excellent quantitative agreement with the corresponding full
$GW$-RPA calculation results in spite of the simplicity of the PPA, establishing the general validity of the
plasmon-pole approximation scheme.  We also provide a comparison between the PPA and the hydrodynamic approximation
in graphene, and comment on the experimental implications of our findings.
\end{abstract}
\maketitle

\section{Introduction}
Graphene many-body effects have been studied extensively, both theoretically and experimentally, for more than
10 years\cite{CastroNetoRMP2009,DasSarmaRMP2011,KotovRMP2012,BasovRMP2014}.  In fact, the theoretical studies\cite{ZhengPRL1996,GonzalezPRL1996,GonzalezPRB2001}
of graphene many-body effects predate the actual laboratory realization of graphene by more than 10 years because
of the fundamental interest in undoped (or intrinsic) graphene being a two-dimensional (2D) nonrelativistic solid
state realization of quantum electrodynamics (QED) with a much larger coupling constant ($\alpha\sim 1$ in graphene,
in contrast to QED where the fine structure constant $\alpha\sim\frac{1}{137}$).  Theories of graphene QED have
matured during the last 15 years\cite{BarnesPRB2014} although the strong-coupling QED aspects of many-body effects
in intrinsic undoped graphene still pose important puzzles.  In particular, why a single-loop weak-coupling perturbative
renormalization group (RG) theory seems to work for undoped graphene with $\alpha\sim 1$ remains a mystery since
the basic perturbation expansion breaks down already at the leading order for $\alpha\sim 1$ in contrast to QED,
where the perturbative expansion is thought to be asymptotic up to 10,000 orders\cite{BarnesPRB2014}.  One possible
reason for the efficacy of a leading-order perturbation theory in the calculation of graphene many-body effects
may be that the $\frac{1}{N}$ type expansion (with $N=2$ for graphene) works here\cite{SonPRB2007,DasSarmaPRB2007}
as has been shown by going to the next-to-leading order in the $\frac{1}{N}$ expansion\cite{HofmannPRL2014} in the theory.
The $\frac{1}{N}$ expansion in graphene is essentially equivalent to the random phase approximation (RPA) of many-body
theory, where the perturbation expansion is carried out in the screened interaction rather than the bare interaction
as in the Hartree-Fock (HF) theory.  Denoting bare and screened interactions formally by $V$ and $W$, respectively,
the leading-order loop expansion (HF theory) and the leading-order $\frac{1}{N}$ expansion (RPA) correspond to $GV$
and $GW$ approximations, respectively, where $W$ is calculated from $V$ by using the random phase approximation for
dynamical screening.  The theory of graphene many-body effects studied as a $GW$-RPA theory, both in undoped intrinsic
and doped extrinsic situations, has been developed and discussed earlier in detail in Refs.\ \onlinecite{DasSarmaPRB2007,HofmannPRL2014,HwangPRB2008,HwangPRB2013,PoliniPRB2008,ProfumoPRB2010}.
We mention that, following Ref.\ \onlinecite{HwangPRL2007}, we define intrinsic (extrinsic) graphene as undoped (doped)
materials with the Fermi level being at the Dirac point (conduction/valence bands for $n$/$p$ doped extrinsic materials).
In this work, we focus on many-body electron-electron interaction effects in doped graphene at zero temperature
as a function of momentum and energy.  This is a problem of experimental relevance since all experiments are typically
carried out in extrinsic graphene although the very low doping density limit may be approaching the intrinsic undoped
limit.  Indeed, there are many experimental reports of the observation of many-body effects in doped graphene\cite{WangNatPhys2012,YuPNAS2013,EliasNatPhys2011,ChaePRL2012,SiegelPRL2013}
that are often compared successfully with $GW$-RPA based theories, and our work should apply to these systems.

In the current theoretical work, we simplify the $GW$-RPA approximation for doped graphene by developing the plasmon-pole
approximation (PPA) for graphene.  PPA is a well-known and extensively used approximation for calculating many-body
effects in Fermi liquids where the electron-electron interaction is via the long-range Coulomb interaction.  Thus, PPA is
a many-body approximation for effective metals, developed originally for simple three-dimensional (3D) metals\cite{LundqvistPKM1967_1,LundqvistPKM1967_2,OverhauserPRB1971}
and later generalized to 2D metals\cite{VinterPRL1975,VinterPRB1976} and 1D metals\cite{DasSarmaPRB1996}.  PPA is an
extensively used approximation for including interaction effects in {\it ab initio} band structure calculations where
self-consistent LDA theories are routinely combined with the $GW$-RPA approximation with the $GW$ part of the calculation
being carried out under PPA rather than in the full RPA\cite{HybertsenPRB1986,NorthrupPRB1989,EngelPRB1993,RohlfingPRB1993,StankovskiPRB2011,CazzanigaPRB2012,LarsonPRB2013,JanssenPRB2015}.
In fact, PPA has been used successfully for studying finite-temperature many-body effects in multi-component Fermi systems
in semiconductor inversion layers\cite{KaliaPRB1978,DasSarmaPRB1979,DasSarmaPRB1982,DasSarmaPRB1983}.  A number of works
have also successfully used {\it ab initio} numerical methods employing PPA in graphene\cite{AttaccalitePSS2009,MakPRL2014}
and in semiconductors. There has, however, been some discussion about the accuracy of various varieties of the PPA as used
in numerical work, including generalizations such as the Hybertsen-Louie and Godby-Needs models, compared to full $GW$ models\cite{StankovskiPRB2011,LarsonPRB2013}.
In addition, a number of numerical codes, such as \textsc{vasp}\cite{ShishkinPRB2006} and \textsc{berkeleygw}\cite{BGW2Website}, can perform
numerical computations in the full $GW$-RPA model as efficiently as PPA-based codes.  It would still be useful, however, to
develop PPA as an analytical approximation for graphene, and in the current work we do exactly
that:  We develop the zero-temperature plasmon-pole approximation to calculate graphene many-body effects within the
standard $GW$-RPA approximation.  The goal here is to develop the analytical graphene PPA theory in detail, emphasizing
several subtle points arising in graphene (but not in ordinary parabolic metals where the PPA has been extensively studied
in the literature), and to explore how well $GW$-PPA duplicates the results of $GW$-RPA self-energy results in graphene,
given the simplicity of PPA as a many-body approximation.  We find that PPA is remarkably effective in doped graphene and $GW$-PPA agrees
quantitatively with $GW$-RPA theories in graphene, and suggest that future many-body calculations in graphene can
safely be carried out in the technically less demanding PPA theories than in the full $GW$-RPA theories, given the
quantitative accuracy of the PPA results we present in this work.  Our work establishes the effectiveness
of PPA independent of the band dispersion or chirality of the system since PPA works as well in graphene with its
linear and chiral energy-momentum dispersion as it does in ordinary non-chiral 2D and 3D metals with parabolic band
dispersions.  Thus, PPA is a quantitatively accurate approximation to the RPA $GW$ self-energy in all metals or doped
semimetals/semiconductors independent of their band dispersion or chirality.

We emphasize, however, that PPA works only for extrinsic graphene with finite doping such that the effective Fermi
energy (in the conduction or valence band depending on whether the doping is $n$ or $p$ type) is much larger than the temperature,
$E_F\gg k_BT$.  Intrinsic (i.e., undoped) graphene has no finite carrier density, and $E_F=0$ (where the energy zero is taken
to be the graphene Dirac point), and PPA is not a meaningful approximation in this gapless situation since the Dirac point is a
quantum critical point separating an electron metal for $E_F>0$ from a hole metal for $E_F<0$.  In particular, the infinite
filled Fermi sea of holes in intrinsic graphene leads to a fundamental problem since the system is now a non-Fermi liquid
by virtue of the Fermi surface being a point (i.e., a Fermi point rather than a Fermi surface).  Extrinsic graphene has a
finite 2D Fermi surface because of doping, and PPA is a meaningful approximation for extrinsic graphene as we show in this
work.  We restrict ourselves to doped graphene with a finite Fermi energy in this work.

The rest of this paper is organized as follows.  In Sec.\ II we develop the basic PPA theory for doped graphene.  In Sec.\
III we provide the numerical results for the graphene self-energy calculated within PPA, comparing the PPA results with
the existing literature on the $GW$-RPA many-body effects.  We conclude in Sec.\ IV with a summary and possible future
directions.  We provide in Appendix A a discussion of the applicability of the $f$-sum rule in graphene, which is closely connected
with the basic formalism of the plasmon pole theory.  In Appendix B we provide a comparison between the hydrodynamic
and plasmon-pole approximations.

\section{Plasmon-pole approximation for doped graphene}
We first develop the PPA formalism for doped graphene.  Before doing so, however, we will first give a brief summary of the existing PPA
for metals\cite{LundqvistPKM1967_1,LundqvistPKM1967_2,OverhauserPRB1971}.  We will specifically consider a 3D metal, although the
PPA is by no means restricted to 3D materials.  In calculating the electron self-energy $\Sigma(\vec{q},\omega)$ within the $GW$
approximation, we obtain
\begin{equation}
\Sigma(\vec{q},\omega)=i\int_{-\infty}^{\infty}\frac{d\nu}{2\pi}\,\int\frac{d^3\vec{k}}{(2\pi)^3}\,G_0(\vec{q}-\vec{k},\omega-\nu)\frac{4\pi e^2}{\kappa k^2}\frac{1}{\epsilon(\vec{k},\nu)},
\end{equation}
where $G_0(\vec{q},\omega)$ is the bare Green's function, $\epsilon(\vec{q},\omega)$ is the dynamical dielectric function,
\begin{equation}
\epsilon(\vec{q},\omega)=1+\frac{4\pi e^2}{\kappa q^2}\Pi(\vec{q},\omega),
\end{equation}
$\kappa$ is the background lattice dielectric constant of the material, and $\Pi(\vec{q},\omega)$ is the electronic polarizability,
whose full form for graphene is shown in Appendix A.  Note that $V(\vec{q})=\frac{4\pi e^2}{\kappa q^2}$ is simply the 3D Coulomb
interaction; in 2D, we would have $V(\vec{q})=\frac{2\pi e^2}{\kappa q}$.  We may split the self-energy into two terms,
\begin{equation}
\Sigma(\vec{q},\omega)=\Sigma_{\text{HF}}(\vec{q},\omega)+\Sigma_{C}(\vec{q},\omega),
\end{equation}
where
\begin{equation}
\Sigma_{\text{HF}}(\vec{q},\omega)=i\int_{-\infty}^{\infty}\frac{d\nu}{2\pi}\,\int\frac{d^3\vec{k}}{(2\pi)^3}\,G_0(\vec{q}-\vec{k},\omega-\nu)\frac{4\pi e^2}{\kappa k^2} \label{Eq:ElectronSE_HF}
\end{equation}
is the Hartree-Fock or exchange self-energy, and
\begin{eqnarray}
\Sigma_{C}(\vec{q},\omega)&=&i\int_{-\infty}^{\infty}\frac{d\nu}{2\pi}\,\int\frac{d^3\vec{k}}{(2\pi)^3}\,G_0(\vec{q}-\vec{k},\omega-\nu)\frac{4\pi e^2}{\kappa k^2} \cr
&\times&\left [\frac{1}{\epsilon(\vec{k},\nu)}-1\right ] \label{Eq:ElectronSE_Corr}
\end{eqnarray}
is the correlation part.  This second term can be difficult to calculate, depending on the form of the dielectric function used.
As a result, the PPA was developed to simplify the calculation of the correlation term\cite{LundqvistPKM1967_1,LundqvistPKM1967_2} $\Sigma_C$.
It consists of replacing the factor in the integrand dependent on the dielectric function with an effective single plasmon mode:
\begin{equation}
\frac{1}{\epsilon(\vec{q},\omega)}-1=\frac{A(\vec{q})}{\pi(\omega^2-\omega_{\vec{q}}^2-i\delta)}, \label{Eq:PPADielectric}
\end{equation}
where $A(\vec{q})$ and $\omega_{\vec{q}}$ are determined from the $f$-sum rule,
\begin{equation}
\int_{0}^{\infty}d\omega\,\omega\mbox{Im}\left [\frac{1}{\epsilon(\vec{q},\omega)}-1\right ]=-\frac{\pi}{2}\omega_p^2, \label{Eq:fSumRule}
\end{equation}
where $\omega_p$ is the long-wavelength plasma frequency, and the zero-frequency Kramers-Kronig relation,
\begin{equation}
\int_{0}^{\infty}d\omega\,\frac{1}{\omega}\mbox{Im}\left [\frac{1}{\epsilon(\vec{q},\omega)}-1\right ]=\frac{\pi}{2}\left [\frac{1}{\epsilon(\vec{q},0)}-1\right ]. \label{Eq:KramersKronig}
\end{equation}
Applying these conditions, one finds that
\begin{eqnarray}
A(\vec{q})&=&\pi\omega_p^2, \label{Eq:EffPlasmon_A} \\
\omega_{\vec{q}}^2&=&-\frac{\omega_p^2}{1/\epsilon(\vec{q},0)-1}. \label{Eq:EffPlasmon_omegaq}
\end{eqnarray}
The correlation term within the PPA is then just
\begin{eqnarray}
\Sigma_{C,\text{PPA}}(\vec{q},\omega)&=&i\int_{-\infty}^{\infty}\frac{d\nu}{2\pi}\,\int\frac{d^3\vec{k}}{(2\pi)^3}\,G_0(\vec{q}-\vec{k},\omega-\nu)\frac{4\pi e^2}{\kappa k^2} \cr
&\times&\frac{\omega_p^2}{\omega^2-\omega_{\vec{q}}^2-i\delta}. \label{Eq:ElectronSE_Corr_PPA}
\end{eqnarray}
The great advantage of PPA is that the integral over frequency in Eq.\ \eqref{Eq:ElectronSE_Corr_PPA} becomes very simple as it is only an
integration over a pole whereas the original $\Sigma_C$ in Eq.\ \eqref{Eq:ElectronSE_Corr} encloses a complicated branch cut arising from
the complicated frequency dependence of the RPA dielectric function.  We show in Appendix A the full form of the RPA dielectric function
$\epsilon(\vec{q},\omega)$ in graphene, emphasizing the complexity of Eq.\ \eqref{Eq:ElectronSE_Corr}.  In metals, this approximation yields
results for the chemical potential differing by only about 1\% from the RPA values\cite{LundqvistPKM1967_2}.  This has led to the extensive
use of PPA in the $GW$ evaluation of metallic self-energy.

Note that Eqs.\ \eqref{Eq:EffPlasmon_A} and \eqref{Eq:EffPlasmon_omegaq} completely fix the functions $\omega_{\vec{q}}$ and
$A(\vec{q})$, the so-called plasmon pole and its strength, respectively, so that all quantities in Eq.\ \eqref{Eq:ElectronSE_Corr_PPA}
are explicitly known, enabling a straightforward evaluation of the self-energy.

If we wish to apply PPA to graphene, however, then we run into a problem.  It turns out that the $f$-sum rule breaks
down for the low-energy effective theory of graphene that we will be employing, as shown formally in a previous work by
two of us\cite{ThrockmortonArXiv} and explicitly demonstrated in Appendix A using the full form of the RPA dielectric
function.  Therefore, strictly speaking, we cannot use the $f$ sum to fix any of the constants appearing in the approximate
expression for $1/\epsilon(\vec{q},\omega)-1$.  We thus wish to determine how one can apply this approximation to graphene.

The Hamiltonian for this system is given by
\begin{equation}
H=v_F\sum_{\vec{r},s}\Psi_s^\dag(\vec{r})\vec{\sigma}\cdot\vec{p}\Psi_s(\vec{r})+\sum_{\vec{r},\vec{r}'}\frac{e^2n(\vec{r})n(\vec{r}')}{\kappa|\vec{r}-\vec{r}'|},
\end{equation}
where the Pauli matrices $\sigma$ act on the sublattice pseudospin, $s$ is the actual spin of the electrons,
$\Psi_s^T(\vec{r})=[a_s(\vec{r}),b_s(\vec{r})]$ is the vector of annihilation operators for the electrons,
$n(\vec{r})=\sum_s\Psi_s^\dag(\vec{r})\Psi_s(\vec{r})$, and $\kappa$ is the dielectric constant of the surrounding
medium.  Here, $v_F\approx 1\times 10^6\text{ m}/\text{s}$ is the graphene Fermi velocity defining the linear
band dispersion $E(\vec{k})=\pm v_F|\vec{q}|$.  Note that we use the effective linear dispersion model, strictly
valid only at low energies, for all energies in the theory.

Since we can only fix one of $A(\vec{q})$ and $\omega_{\vec{q}}$ using the Kramers-Kronig relation, we will fix
$A(\vec{q})$ and consider three different models for $\omega_{\vec{q}}$:

1) The static RPA (SRPA) model, which was introduced by Vinter\cite{VinterPRL1975,VinterPRB1976} for 2D metals,
\begin{equation}
\omega_{\vec{q}}^2=-\frac{\omega_p^2}{1/\epsilon(\vec{q},0)-1}, \label{Eq:PPA_StaticRPA}
\end{equation}
where $\omega_p=\omega_0\sqrt{q}$, $\omega_0=\sqrt{\frac{2e^2v_F\sqrt{\pi n}}{\kappa}}$, and $n$ is the electron
number density.  Here, for $\epsilon(\vec{q},0)$, we use the (exact) RPA dielectric function.  Note that $\omega_p$
is the long wavelength plasma frequency for graphene. 

2) The Thomas-Fermi (TF) model,
\begin{equation}
\omega_{\vec{q}}^2=\omega_p^2\left (1+\frac{q}{k_{\text{TF}}}\right ), \label{Eq:PPA_ThomasFermi}
\end{equation}
where
\begin{equation}
k_{\text{TF}}=\frac{4e^2\sqrt{\pi n}}{\kappa v_F} \label{Eq:TFWaveNumber}
\end{equation}
is the Thomas-Fermi screening wave number.

3) The hydrodynamic (HD) model,
\begin{equation}
\omega_{\vec{q}}^2=\omega_p^2+v_F^2 q^2. \label{Eq:PPA_Hydrodynamic}
\end{equation}

We now motivate these three approximations.  In the standard 3D and 2D PPA for parabolically dispersing metals
(with no Dirac point by definition), the PPA is motivated by the fact that the effective plasmon-pole frequency $\omega_{\vec{q}}$
defining the effective dielectric function, Eq.\ \eqref{Eq:PPADielectric}, should behave as the long-wavelength
plasma frequency $\omega_p$ and the single-particle energy dispersion $\frac{q^2}{2m}$, respectively, in the long-wavelength
(i.e., $q\to 0$) and short-wavelength (i.e., $q\to\infty$) limits.  This is, in fact, guaranteed by Eqs.\ \eqref{Eq:fSumRule}--\eqref{Eq:EffPlasmon_omegaq}
combined with Eq.\ \eqref{Eq:ElectronSE_Corr} for a parabolically dispersing electron energy band, where the PPA has
so far been used.  This does not, however, happen for 2D graphene as discussed below.

Using Eq.\ \eqref{Eq:EffPlasmon_A} in Eq.\ \eqref{Eq:PPADielectric}, we get
\begin{equation}
\epsilon_{\text{PPA}}(\vec{q},\omega)=\frac{\omega^2-\omega_{\vec{q}}^2}{\omega^2-\omega_{\vec{q}}^2+\omega_p^2}, \label{Eq:DielectricFunc_PPA}
\end{equation}
leading to the $\epsilon_{\text{PPA}}(\vec{q},\omega)=0$ simple pole condition being given by $\omega=\omega_{\vec{q}}$.
Using Eq.\ \eqref{Eq:EffPlasmon_omegaq}, we get
\begin{equation}
\omega_{\vec{q}}^2=-\frac{\omega_p^2}{1/\epsilon(\vec{q},0)-1}=-\frac{\epsilon(\vec{q},0)\omega_p^2}{1-\epsilon(\vec{q},0)}, \label{Eq:PlasmonPole_Def}
\end{equation}
which, when combined with the RPA form for the graphene static dielectric function, leads to
\begin{equation}
\omega_{\vec{q}}\sim\sqrt{q} \label{Eq:PlasmonPole_SRPA_Extreme_q}
\end{equation}
for both the $q\to 0$ and $q\to\infty$ limits.  In obtaining Eq.\ \eqref{Eq:PlasmonPole_SRPA_Extreme_q}, we have used the exact
form for $\epsilon_{\text{RPA}}(\vec{q})$ as given in Ref.\ \onlinecite{HwangPRB2007} (see Appendix A):
\begin{equation}
\epsilon_{\text{RPA}}(\vec{q})=1+\frac{2\pi e^2}{\kappa q}\Pi_0(\vec{q}),
\end{equation}
with
\begin{eqnarray}
\Pi_0(\vec{q})&=&D(E_F)\left\{1+\frac{\pi q}{8k_F}-\theta(2k_F-q)\frac{\pi q}{8k_F}\right. \cr
&-&\left.\theta(q-2k_F)\left [\tfrac{1}{2}\sqrt{1-\frac{4k_F^2}{q^2}}+\frac{q}{4k_F}\arcsin\left (\frac{2k_F}{q}\right )\right ]\right\}, \nonumber \\
\end{eqnarray}
where
\begin{equation}
D(E_F)=\frac{2k_F}{\pi v_F}
\end{equation}
is the graphene density of states, and $k_F=\sqrt{\pi n}$ is the Fermi wave number.  Thus, the incorporation of the static
RPA dielectric function into the PPA leads to an effective plasmon-pole frequency $\omega_{\vec{q}}$ that behaves as the
long-wavelength plasma frequency $\sim\sqrt{q}$ both in the $q\to 0$ and $q\to\infty$ limits in contrast to the corresponding
parabolic PPA.  This is the approximation defined by Eq.\ \eqref{Eq:PPA_StaticRPA} above.

This ``problem'' is, however, fixed by using the Thomas-Fermi dielectric function $\epsilon_{\text{TF}}(\vec{q},0)$ instead of the
exact static RPA dielectric function in Eq.\ \eqref{Eq:EffPlasmon_omegaq} for $\omega_{\vec{q}}$.  The Thomas-Fermi dielectric
function is simply the long wavelength limit of $\epsilon_{\text{RPA}}(\vec{q})$:
\begin{equation}
\epsilon_{\text{TF}}(\vec{q})\equiv\epsilon_{\text{RPA}}(q\to 0),
\end{equation}
leading to
\begin{equation}
\epsilon_{\text{TF}}(\vec{q})=1+\frac{2\pi e^2}{\kappa q}D(E_F)=1+\frac{k_{\text{TF}}}{q}, \label{Eq:TFDielectric}
\end{equation}
where $k_{\text{TF}}=\frac{e^2}{2\kappa}D(E_F)=\frac{4e^2\sqrt{\pi n}}{\kappa v_F}$ is the Thomas-Fermi screening wave number defined
in Eqs.\ \eqref{Eq:PPA_ThomasFermi} and \eqref{Eq:TFWaveNumber} above.  Putting Eq.\ \eqref{Eq:TFDielectric} for $\epsilon(\vec{q},0)$
into Eq.\ \eqref{Eq:PlasmonPole_Def}, we get
\begin{equation}
\omega_{\vec{q}}\sim\begin{cases}
\sqrt{q}, & q\to 0 \\
v_F q, & q\to\infty
\end{cases} \label{Eq:PlasmonPole_TF_Limits}
\end{equation}
Note that Eq.\ \eqref{Eq:PlasmonPole_TF_Limits} does provide the asymptotic forms for $\omega_{\vec{q}}$ going as the plasma frequency
and the single-particle frequency in the long- and short-wavelength limits, respectively.

Finally, the hydrodynamic approximation, Eq.\ \eqref{Eq:PPA_Hydrodynamic}, assumes the following effective hydrodynamic dielectric
function:
\begin{equation}
\epsilon_{\text{HD}}(\vec{q},\omega)=1-\frac{\omega_p^2}{\omega^2-v_F^2 q^2},
\end{equation}
which then leads to the following effective hydrodynamic plasma frequency (see Eq.\ \eqref{Eq:PPA_Hydrodynamic} above):
\begin{equation}
\omega_{\text{HD}}^2(\vec{q})=\omega_p^2+v_F^2 q^2,
\end{equation}
where $\omega_{\text{HD}}(\vec{q})$ is the solution to the usual $\epsilon_{\text{HD}}(\vec{q},\omega)$ for a collective mode.
Letting $\omega_{\vec{q}}=\omega_{\text{HD}}(\vec{q})$ gives
\begin{equation}
\omega_{\vec{q}}^2=\omega_p^2+v_F^2 q^2\equiv\omega_{\text{HD}}^2(\vec{q}),
\end{equation}
as in Eq.\ \eqref{Eq:PPA_Hydrodynamic} above, leading to a hydrodynamic PPA dielectric function.  Using Eqs.\ \eqref{Eq:PPADielectric}
and \eqref{Eq:EffPlasmon_A}, we get
\begin{equation}
\frac{1}{\epsilon_{\text{PPA}}^{\text{HD}}(\vec{q},\omega)}=1+\frac{\omega_p^2}{\omega^2-\omega_{\vec{q}}^2},
\end{equation}
or
\begin{equation}
\epsilon_{\text{PPA}}^{\text{HD}}(\vec{q},\omega)=\frac{\omega^2-\omega_{\vec{q}}}{\omega^2-\omega_{\vec{q}}^2+\omega_p^2}. \label{Eq:DielectricFunc_Hyd}
\end{equation}

\begin{figure}[tb]
	\centering
		\includegraphics[width=\columnwidth]{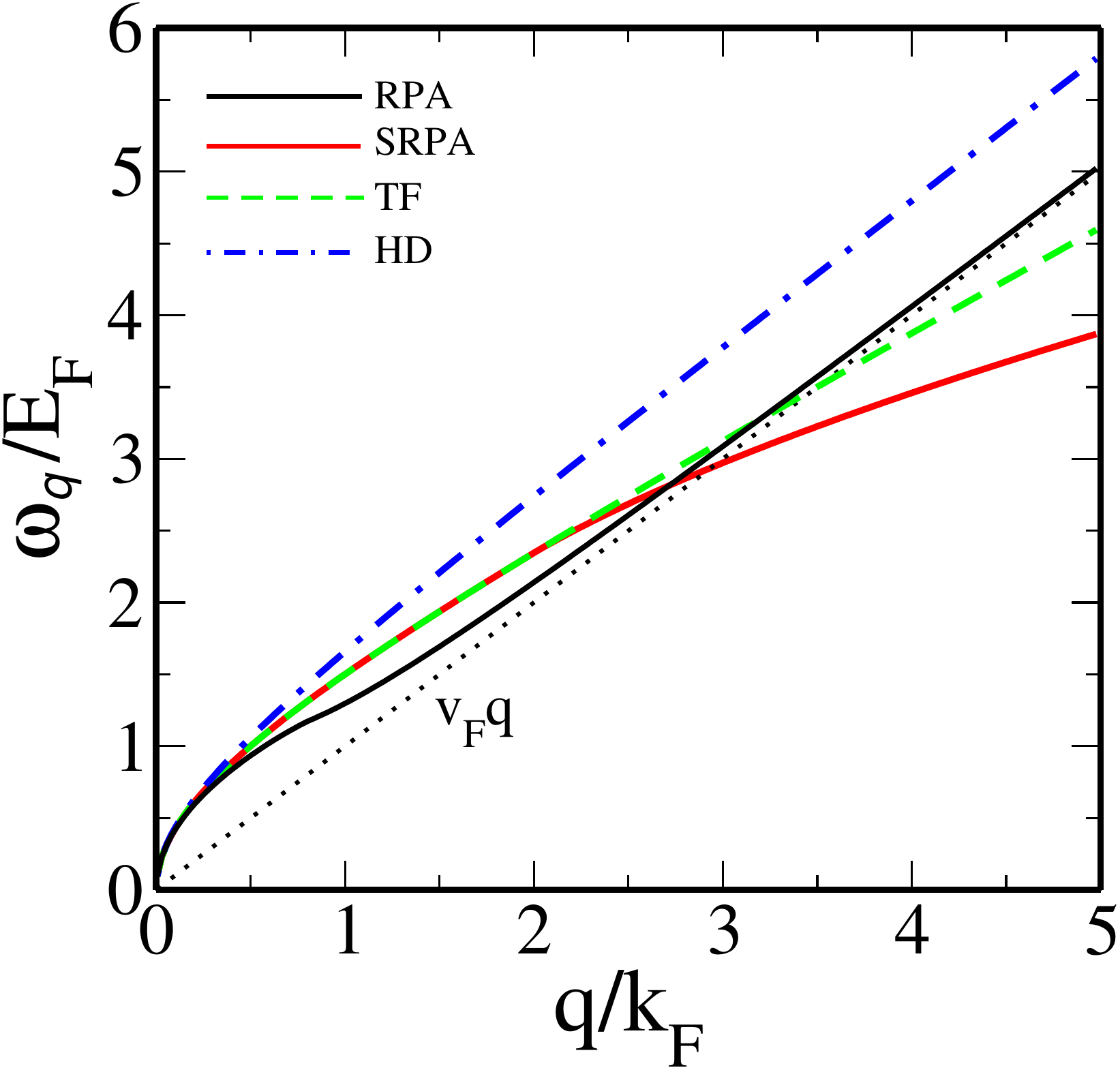}
	\caption{Plots of the plasmon dispersion as a function of wave vector from the full RPA dielectric function and from the static RPA (SRPA), Thomas-Fermi (TF), and hydrodynamic (HD) models.}
	\label{Fig:PlasmonDisp}
\end{figure}

We note that these three approximations, all defined through the function $\omega_{\vec{q}}$ as in Eqs.\ \eqref{Eq:PPA_StaticRPA}--\eqref{Eq:PPA_Hydrodynamic},
provide three different plasmon pole approximations for the dynamical PPA dielectric function,
\begin{equation}
\epsilon_{\text{PPA}}(\vec{q},\omega)=\frac{\omega^2-\omega_{\vec{q}}^2}{\omega^2-\omega_{\vec{q}}^2+\omega_p^2},
\end{equation}
since, although $\omega_p^2=\frac{2e^2v_Fq\sqrt{\pi n}}{\kappa}$, the long-wavelength graphene plasma frequency, is the same in all
three approximations (static RPA, Thomas-Fermi, hydrodynamic), the effective plasmon pole frequency $\omega_{\vec{q}}$, defined by
$\epsilon_{\text{PPA}}(\vec{q},\omega_{\vec{q}})=0$, is diffferent in the three schemes.  The effective $\omega_{\vec{q}}$ is shown in
Fig.\ \ref{Fig:PlasmonDisp} as a function of $q$ for the three approximations compared with the exact RPA result.

As we will show, the PPA self-energies obtained from the three approximations are very similar in magnitude, all agreeing well with
the RPA self-energy results, thus well justifying our plasmon pole approximation scheme in graphene independent of the precise form
of the approximation used in the theory.
We note that the PPA involves a pure pole (at $\omega=\omega_{\vec{q}}$) approximation for the dynamical dielectric function $\epsilon(\vec{q},\omega)$
instead of the much more complex pole and branch cut form for $\epsilon(\vec{q},\omega)$ in the full RPA\cite{HwangPRB2007}.  Using
Eq.\ \eqref{Eq:ElectronSE_Corr_PPA} and the form of $\epsilon(\vec{q},\omega)$ in PPA, we can calculate the imaginary part of the PPA
self-energy by doing the frequency $\nu$ integration in Eq.\ \eqref{Eq:ElectronSE_Corr_PPA} analytically and then reducing the 2D
momentum integral over $\vec{k}$ to a simple 1D integral over the magnitude $k$.  Then, the real part of the self-energy is easily
obtained by using the Kramers-Kronig relation involving one more frequency integration.  This simplification of the self-energy into
simple real integrals, only a 1D real integral for $\mbox{Im }\Sigma(\vec{q},\omega)$ and a 2D real integral for $\mbox{Re }\Sigma(\vec{q},\omega)$, is
what makes the PPA attractive (and much less computationally demanding) compared with the full RPA.

In Appendix B, we provide a comparative discussion between the hydrodynamic and plasmon-pole approximations to the dielectric function
since they share the superficial similarity of having just a simple pole describing the frequency response.

\section{Numerical results}

We now present our numerical results.  We plot the plasmon dispersion relations obtained from the full RPA dielectric function and
from the three models that we just presented in the context of our plasmon-pole approximation [Eqs.\ \eqref{Eq:PPA_StaticRPA}--\eqref{Eq:PPA_Hydrodynamic}]
in Fig.\ \ref{Fig:PlasmonDisp}.  We assume throughout that $\kappa=2.4$, corresponding to graphene on a SiO${}_2$ substrate, and
$n=10^{12}\text{ cm}^{-2}$.  Of course, the qualitative results and the relative validity of PPA compared with RPA are independent
of the choice of density and background dielectric constant.  It was demonstrated by two of us\cite{HwangPRB2007} that the full RPA
dielectric function yields a plasmon frequency that is proportional to $\sqrt{q}$ for $q\ll k_F$ and to $q$ for $q\gg k_F$.  The static
RPA model yields a plasmon frequency that is strictly proportional to $\sqrt{q}$, i.e., it captures the correct low-energy behavior,
but fails to capture the proper high-energy dependence.  The TF model gives the correct behavior for both the low- and high-energy
limits, but yields the wrong coefficient for the high-energy case.  The hydrodynamic model also gives the correct dependence in both
limits, and even yields the correct coefficient in the high-energy case, only being offset from the full RPA result by a constant of
$\alpha=e^2/(\kappa v_F)$ corresponding to the coupling constant in graphene. This indicates that for a small coupling constant ($\alpha\ll 1$)
the hydrodynamic model gives the plasmon dispersion predicted by full RPA calculation.  Thus, the effective plasmon pole frequency
$\omega_{\vec{q}}$ varies among the three approximation schemes, all of them differing somewhat from RPA.  Since the $GW$ approximation
itself is likely to be a good approximation only for $\alpha$ not too large, one can safely use the hydrodynamic PPA approximation in
carrying out self-energy calculations for doped graphene.  Note that, unlike parabolic metals, the linear dispersion in graphene with
a constant Fermi velocity makes the coupling constant $\alpha$ independent of carrier density or doping.

\begin{figure*}[tb]
	\centering
		\includegraphics[width=2\columnwidth]{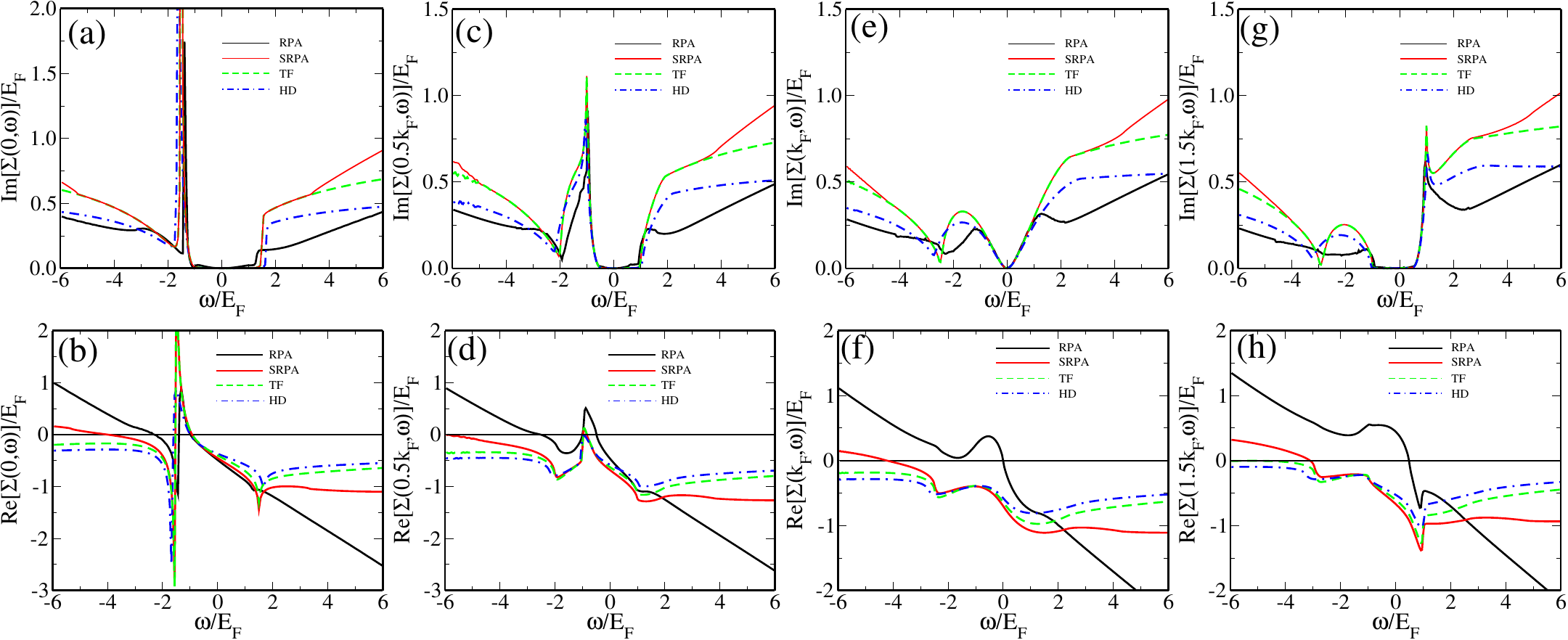}
	\caption{Plots of the imaginary (top) and real (bottom) parts of the electron self-energy $\Sigma(\vec{q},\omega)$ within various
	approximations as functions of frequency for $q=0$ [(a) and (b)], $q=0.5k_F$ [(c) and (d)], $q=k_F$ [(e) and (f)], and $q=1.5k_F$ [(g) and (h)].}
	\label{Fig:SelfEnergy}
\end{figure*}

We now turn our attention to the electron self-energy $\Sigma(\vec{q},\omega)$.  We provide plots of the calculated self-energy for
different values of $k$ as functions of frequency in Fig.\ \ref{Fig:SelfEnergy} for both the $GW$-RPA
and for the PPA with the three plasmon dispersions given earlier (see Fig.\ \ref{Fig:PlasmonDisp}).  Here, we consider $q=0$, $0.5k_F$,
$k_F$, and $1.5k_F$.  We note that all four approximations agree very well with each other for the imaginary part of the self-energy
for small frequencies.  However, for large $\omega$ the real part of self-energy from PPA is qualitatively different from that of RPA.
The RPA predicts that the real part of the self-energy increases linearly with $\omega$, but all PPA results show that the real part of
the self-energy saturates in this region.  This linear increase of the real part of self-energy arises from the single-particle (electron-hole)
excitation contribution, which is absent in PPA.  However, the important structures (deep or step increase) in the self-energy arising
from the coupling of plasmon absorption or emission (or plasmaron production) agree well in all approximations. The disagreement between
RPA and PPA results at large $\omega$ (or large off-shell energy) is not important in the quasiparticle properties of graphene because
the spectral function weight of the quasiparticles decreases with increasing off-shell energy.  This indicates that the PPA, regardless of
the specific model for $\omega_{\vec{q}}$, should reliably predict the quasiparticle spectrum.  We emphasize that the differences with
RPA are all quantitative and not qualitative.

\begin{figure*}[tb]
	\centering
		\includegraphics[width=2\columnwidth]{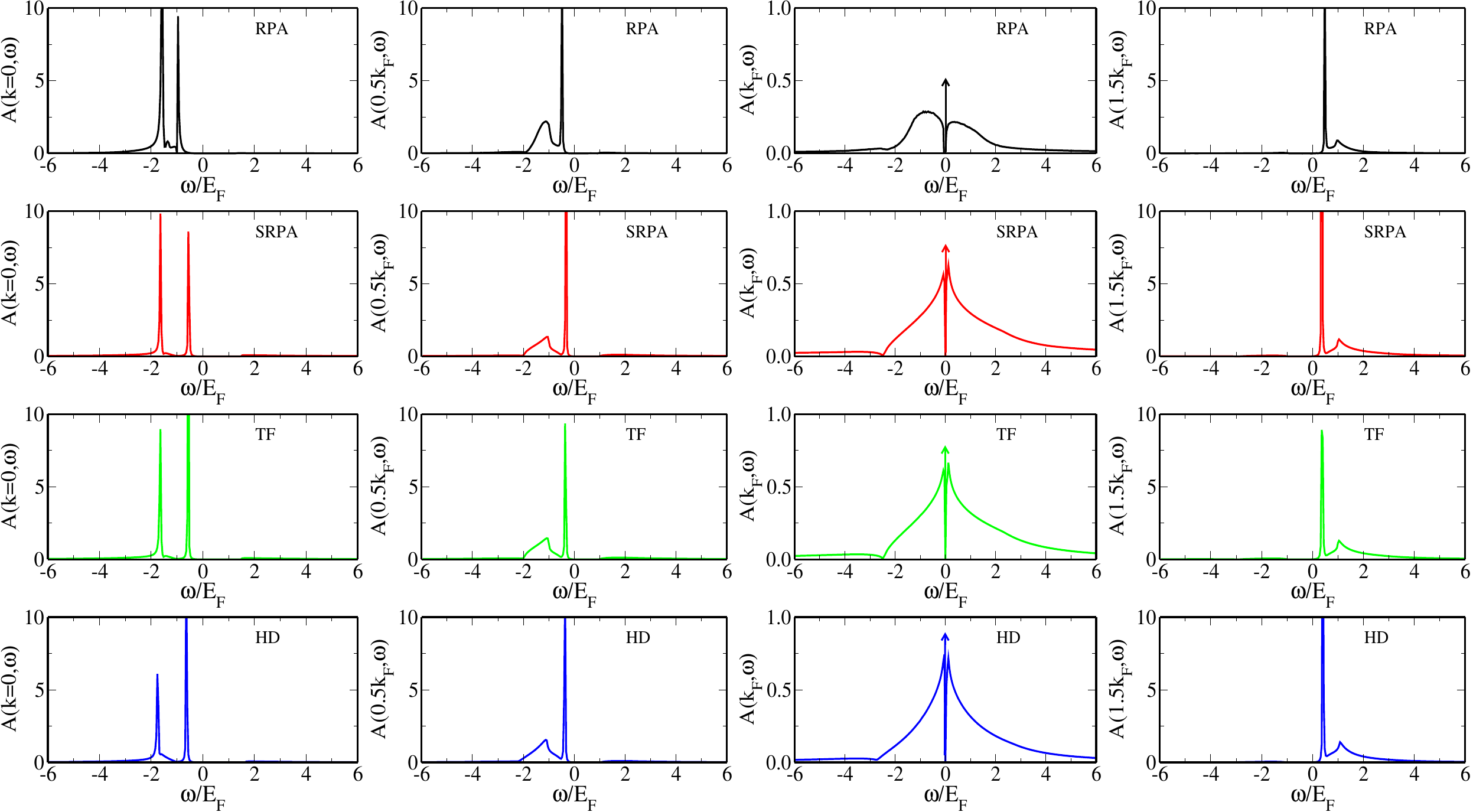}
	\caption{Plots of the spectral function $A(\vec{q},\omega)$ within various approximations as functions of frequency for (from
	left to right) $q=0$, $0.5k_F$, $k_F$, and $1.5k_F$.}
	\label{Fig:SpectralFunction}
\end{figure*}

We also provide plots of the spectral function, $A(\vec{q},\omega)$, in Fig.\ \ref{Fig:SpectralFunction}.  Once the self-energy
$\Sigma(\vec{q},\omega)$ is known, the single-particle spectral function $A(\vec{q},\omega)$ can be calculated. The spectral function
contains important dynamical information about the system and is given by
\begin{equation}
A(\vec{q},\omega) = \frac{2\mbox{Im}\Sigma(\vec{q},\omega)}{\left [ \omega - E_0(\vec{q}) - \mbox{Re}\Sigma(\vec{q},\omega) \right ]^2 + \left [ \mbox{Im} \Sigma(\vec{q},\omega) \right ]^2 },
\end{equation}
where $E_0(\vec{q}) = v_F q - E_F$ is the single particle energy measured from the Fermi energy.  The spectral function, $A(\vec{q},\omega)=-2\mbox{Im}G(\vec{q},\omega)$, is
simply the imaginary part of the interacting Green's function, indicating the spectral weight of the system in the $(\vec{q},\omega)$ space. The non-interacting spectral function is
a $\delta$ function at the noninteracting energy $v_Fq$, but in the presence of interaction effects, the finite value of the imaginary part of self-energy Im$\Sigma(\vec{q},\omega) \neq 0$
broadens the single particle $\delta$-function peak except at $q=k_F$ and $\omega = E_F$, where Im$\Sigma =0$.  Note that the chemical potential
of the interacting electron gas is determined by setting $q=k_F$ and $\omega=0$ in the above equation to guarantee a non-zero Fermi surface.  As
expected, we find good qualitative agreement among all of the approximations except at $q=k_F$, for which the spectral function behaves differently
near the delta function singularity for the $GW$-RPA compared with the PPA.  For $q\neq k_F$, all four approximations predict two peaks in the spectral
function, corresponding to two excitations.  The quasiparticle peaks occur at $\omega = E_0(\vec{q}) - E_F$. The other peak is known as a ``plasmaron''
mode.  These results are also very well known in 2D and 3D metals\cite{LundqvistPKM1967_2}.  We see this plasmaron peak for nonzero $q$ as well,
though it is much broadened.  For $q<k_F$, the plasmaron peak appears below the quasiparticle peak.  At $q=k_F$ the plasmaron peak does not appear
at all, and for $q>k_F$ the plasmaron peak appears above the quasiparticle peak.  For $q=0$, the plasmaron peak appears around $\omega\approx -1.8E_F$.
In RPA the plasmaron peak has larger spectral weight than that of the quasiparticle peak, but for PPA the quasiparticle peak becomes smaller than
that of plasmaron.  This trend seems true for $q<k_F$. However, for $q>k_F$ the behavior is reversed.  We should point out, however, that,
in more refined approximations, such as cumulant expansions of the Green's functions\cite{LischnerPRL2013,LischnerPRB2013}, these ``plasmaron'' peaks
are not as pronounced as they are in our results.  The issue of the existence or not of true plasmaron peaks in electronic spectral function therefore
has remained somewhat controversial.  We should, however, point out that experiments\cite{BostwickScience2010,WalterPRB2011} claim to have
seen such peaks in graphene, and indeed standard $GW$-RPA theories in graphene produce well-defined plasmaron peaks\cite{HwangPRB2008,HwangPRB2013,PoliniPRB2008}.
Our work is, however, not aimed at interpreting experimental data or establishing whether or not plasmaron peaks exist; we only seek to show
that PPA is essentially as good a many-body approximation in graphene as RPA itself is, which is manifestly obvious from our Figs.\ \ref{Fig:SelfEnergy} and \ref{Fig:SpectralFunction}.

The fact that the calculated PPA spectral function (essentially in all three of our PPA schemes) agrees well with the RPA result is significant since
the spectral function determines the quasiparticle properties as observed in ARPES\cite{SiegelPRL2013} and STM\cite{ChaePRL2012} measurements.  This
good agreement for the calculated graphene spectral function between PPA and RPA indicates that PPA should be a good approximation for calculating
graphene many-body effects in future theoretical works.  Since PPA, with its single-pole description of electronic carrier response, is substantially
easier to implement computationally than RPA, our explicit validation of PPA with respect to the calculated spectral function becomes particularly
useful.  In this context, we emphasize that we find that all three PPA schemes, i.e., static RPA [Eq.\ \eqref{Eq:PPA_StaticRPA}], the Thomas-Fermi approximation
[Eq.\ \eqref{Eq:PPA_ThomasFermi}], and the hydrodynamic approximation [Eq.\ \eqref{Eq:PPA_Hydrodynamic}], work equally well, and hence any of them
should be suitable for future theoretical works on graphene many-body effects.  Since the hydrodynamic approximation [Eq.\ \eqref{Eq:PPA_Hydrodynamic}]
is the simplest one among the three we consider, we recommend the use of the hydrodynamic PPA [i.e., Eq.\ \eqref{Eq:PPA_Hydrodynamic}] for future
theoretical calculations (see Appendix B in this context).  It is also in some sense the ``best'' approximation since it reproduces well the RPA plasmon
dispersion (Fig.\ \ref{Fig:PlasmonDisp}).

Finally, we comment on the well-known ``running of the coupling constant'' issue in graphene many-body effects\cite{GonzalezPRL1996,GonzalezPRB2001,BarnesPRB2014,SonPRB2007,DasSarmaPRB2007,HofmannPRL2014}.
Since we are considering doped extrinsic graphene with a fixed density, the ultraviolet divergence associated with the chiral linear graphene dispersion
is simply a weak logarithmic correction going as $1+\frac{\alpha}{4}\log\left (\frac{k_c}{k_F}\right )$, where $\alpha=\frac{e^2}{\kappa v_F}$ and $k_c$,
the ultraviolet cutoff, is chosen in our calculation to be $\frac{1}{a}$ where $a=0.246\text{ nm}$ is the graphene lattice constant.  We choose $v_F=1\times 10^6\,\text{m}/\text{s}$
throughout in our calculation as the graphene Fermi velocity.  Since the Fermi wave number $k_F=\sqrt{\pi n}$ is fixed at a fixed carrier density, the
ultraviolet divergence simply represents a constant (and small) shift in the graphene self-energy which varies slightly as the density changes.  In our
theory, this logarithmic self-energy correction arising from the high momentum cutoff $k_c$ is absorbed entirely into the exchange self-energy or the HF
part [i.e., Eq.\ \eqref{Eq:ElectronSE_HF}] as discussed in detail in Refs.\ \onlinecite{DasSarmaPRB2007} and \onlinecite{HwangPRL2007}.  Since the PPA
deals with the infrared divergence arising from the long-range Coulomb interaction in the correlation self-energy of Eq.\ \eqref{Eq:ElectronSE_Corr}, the
ultraviolet divergence of the running of the coupling constant does not pose any additional problem in the context of using RPA.  Thus, the ultraviolet divergence
is already a problem with intrinsic graphene (i.e., no doping) which we regularize by having a lattice cutoff whereas our PPA then deals with the dynamical
response of the doped carriers present in the experimental doped samples.  The main goal of the current work is obtaining a good approximation for the wave
number and the frequency dependence of the graphene spectral function at a fixed doping level (i.e., fixed Fermi energy or wave vector), and as such the
ultraviolet running of the coupling constant is not germane in our consideration.  More details on this topic may be found in Refs.\ \onlinecite{DasSarmaPRB2007},
\onlinecite{HofmannPRL2014}, \onlinecite{HwangPRL2007}, and particularly in Ref.\ \onlinecite{DasSarmaArXiv}.  We emphasize that the logarithmic
corrections arising from the ultraviolet cutoff are fully included in our PPA theories, but they have no qualitative effect in determining the
momentum- and energy-dependent spectral function at a fixed carrier density.

\section{Conclusion}
We have developed the plasmon-pole approximation for calculating the electron-electron interaction-induced many-body effects in the spectral
function of doped or extrinsic graphene.  Since the single-band effective chiral linear dispersion model for graphene does not obey the simple
$f$-sum rule by virtue of the infinite filled Fermi sea in the valence band\cite{ThrockmortonArXiv}, the PPA is not unique as it is in 3D\cite{LundqvistPKM1967_1,LundqvistPKM1967_2,OverhauserPRB1971}
or 2D\cite{VinterPRL1975,VinterPRB1976,DasSarmaPRB1996} metals.  We introduced three distinct approximations for obtaining the effective plasmon-pole
frequency using static RPA, the Thomas-Fermi approximation, and the hydrodynamic dielectric function, respectively.  It turns out that all
three PPA schemes, as we show through explicit calculations, give many-body renormalization, specifically the interacting spectral function,
very similar to that obtained with the full $GW$-RPA theory, thus validating all three approximation schemes more or less equivalently.  Given
the simplicity of the hydrodynamic PPA, as defined by Eq.\ \eqref{Eq:PPA_Hydrodynamic} for the effective plasmon-pole frequency, we suggest
that future graphene many-body calculations utilize the hydrodynamic PPA introduced in this work.

Possible future generalizations of our work could involve the development of the PPA for 3D Dirac-Weyl materials where the collective plasmon
response has been experimentally observed\cite{SushkovPRB2015}.  We believe that PPA should be valid in 3D Dirac systems as well, but obviously
a 3D generalization of our work is necessary for a definitive conclusion.  Another possible application of our theory could be the development
of PPA for bilayer graphene with its approximately parabolic band dispersion\cite{SensarmaPRB2011} where a comparison with the existing RPA
many-body results could validate (or not) the plasmon-pole approximation in bilayer graphene.  There is no {\it a priori} reason to expect that
a PPA similar to that described in this work cannot be developed for other systems (e.g., bilayer graphene); however, any such theory would need
to be validated by showing that it produces results sufficiently close to those produced by, say, a full RPA calculation.

\acknowledgments
This work is supported by the Laboratory for Physical Sciences.

\appendix
\section{$f$-sum rule in graphene}
Here, we will attempt to calculate the $f$ sum for the polarizability of graphene within
RPA.  The $f$-sum rule is given by
\begin{equation}
F=\int_{0}^{\infty}d\omega\,\omega\,\mbox{Im}\left [\frac{1}{\epsilon(\vec{q},\omega)}-1\right ]. \label{Eq:FSumRule}
\end{equation}
As stated in the main text, the usual $f$-sum rule states that this integral should evaluate to $-\frac{\pi}{2}\omega_p^2$,
where $\omega_p$ is the low-wavelength plasmon dispersion\cite{NozieresPR1957}.  In a previous work by two of us\cite{ThrockmortonArXiv},
we formally showed that this rule breaks down when a second, negative-energy, infinitely filled valence band is present, as
is the case in the low-energy effective theory of graphene employed in this work; here, we will demonstrate this
breakdown explicitly.  The dielectric function $\epsilon(\vec{q},\omega)=1+V(\vec{q})\Pi(\vec{q},\omega)$, where
$V(\vec{q})=\frac{2\pi e^2}{\kappa q}$ and $\Pi(\vec{q},\omega)$ is the polarizability.  We will use the RPA
expression found in Ref.\ \onlinecite{HwangPRB2007}, which we state here for convenience.  It is independent of
the direction of $\vec{q}$, so we will write the dielectric function as $\epsilon(q,\omega)$ and the polarizability
as $\Pi(q,\omega)$ from this point forward.  If we define $x=q/k_F$, $\nu=\omega/E_F$, and $\tilde{\Pi}(x,\nu)=\Pi(q,\omega)/D_0$,
where $D_0=\frac{Nk_F}{2\pi v_F}$ is the density of states at the Fermi energy and $N$ is the number of Dirac
cones (4 for graphene, including spin and valley), then the polarizability may be split up into two contributions, $\Pi^{+}$ and $\Pi^{-}$:
\begin{equation}
\tilde{\Pi}(x,\nu)=\tilde{\Pi}^{+}(x,\nu)+\tilde{\Pi}^{-}(x,\nu).
\end{equation}
The term, $\Pi^{+}$, divides further into
\begin{equation}
\tilde{\Pi}^{+}(x,\nu)=\tilde{\Pi}_1^{+}(x,\nu)\theta(\nu-x)+\tilde{\Pi}_2^{+}(x,\nu)\theta(x-\nu).
\end{equation}
The real and imaginary parts of $\tilde{\Pi}_{1,2}^{+}$ are then given by
\begin{eqnarray}
\mbox{Re}\,\tilde{\Pi}_1^{+}(x,\nu)&=&1-\frac{1}{8\sqrt{\nu^2-x^2}}\{f_1(x,\nu)\theta(\left |2+\nu\right |-x) \cr
&+&\mbox{sgn}(\nu-2+x)f_1(x,-\nu)\theta(\left |2-\nu\right |-x)+\cr
&+&f_2(x,\nu)[\theta(x+2-\nu)+\theta(2-x-\nu)]\}, \nonumber \\ \\
\mbox{Im}\,\tilde{\Pi}_1^{+}(x,\nu)&=&-\frac{1}{8\sqrt{\nu^2-x^2}}\{f_3(x,-\nu)\theta(x-\left |\nu-2\right |) \cr
&+&\tfrac{1}{2}\pi x^2[\theta(x+2-\nu)+\theta(2-x-\nu)]\}, \\
\mbox{Re}\,\tilde{\Pi}_2^{+}(x,\nu)&=&1-\frac{1}{8\sqrt{\nu^2-x^2}}\{f_3(x,\nu)\theta(x-\left |\nu+2\right |) \cr
&+&f_3(x,-\nu)\theta(x-\left |\nu-2\right |)+\cr
&+&\tfrac{1}{2}\pi x^2[\theta(\left |\nu+2\right |-x)+\theta(\left |\nu-2\right |-x)]\}, \nonumber \\ \\
\mbox{Im}\,\tilde{\Pi}_2^{+}(x,\nu)&=&\frac{\theta(\nu-x+2)}{8\sqrt{\nu^2-x^2}}[f_4(x,\nu) \cr
&+&f_4(x,-\nu)\theta(2-x-\nu)],
\end{eqnarray}
where the functions $f_i(x,\nu)$ are
\begin{eqnarray}
f_1(x,\nu)&=&(2+\nu)\sqrt{(2+\nu)^2-x^2} \cr
&-&x^2\ln\left [\frac{\sqrt{(2+\nu)^2-x^2}+\nu+2}{\left |\sqrt{\nu^2-x^2}+\nu\right |}\right ], \\
f_2(x,\nu)&=&x^2\ln\left (\frac{\nu-\sqrt{\nu^2-x^2}}{x}\right ), \\
f_3(x,\nu)&=&(2+\nu)\sqrt{(2+\nu)^2-x^2}+x^2\arcsin\left (\frac{2+\nu}{x}\right ), \nonumber \\ \\
f_4(x,\nu)&=&(2+\nu)\sqrt{(2+\nu)^2-x^2} \cr
&-&x^2\ln\left [\frac{\sqrt{(2+\nu)^2-x^2}+\nu+2}{x}\right ].
\end{eqnarray}
$\Pi^{-}$, on the other hand, is simply given by
\begin{equation}
\tilde{\Pi}^{-}(x,\nu)=\frac{\pi x^2}{8\sqrt{x^2-\nu^2}}\theta(x-\nu)+i\frac{\pi x^2}{8\sqrt{\nu^2-x^2}}\theta(\nu-x).
\end{equation}

The dielectric function can be rewritten in terms of $\tilde{\Pi}(x,\nu)$ as follows:
\begin{equation}
\epsilon(x,\nu)=1+\frac{N\alpha}{x}\tilde{\Pi}(x,\nu),
\end{equation}
where $\alpha=\frac{e^2}{\kappa v_F}$ is the effective fine structure constant.  This is approximately
$\frac{2.2}{\kappa}$ for graphene.

\begin{figure}[htb]
	\centering
		\includegraphics[width=\columnwidth]{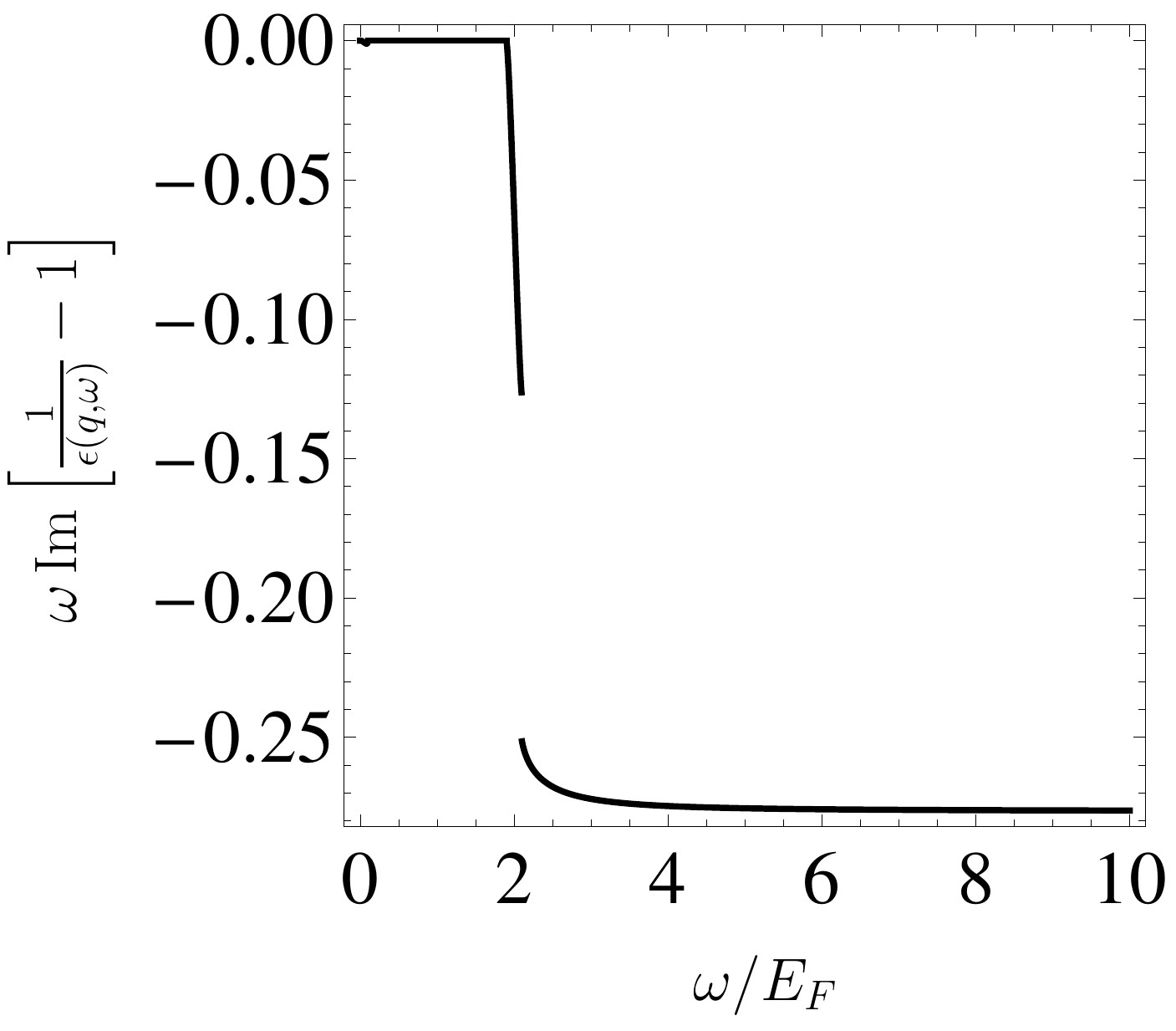}
		\caption{Plot of the integrand in the $f$-sum rule, Eq.\ \eqref{Eq:FSumRule}, as a function of $\omega$.
		Here, $q=0.1k_F$ and $\alpha\approx 0.9$.}
	\label{Fig:fSumIntegrand}
\end{figure}

We now provide a plot of the integrand in Eq.\ \eqref{Eq:FSumRule} for $q=0.1k_F$ in Fig.\ \ref{Fig:fSumIntegrand}.
In our numerical work we take $\alpha\approx 0.9$ as appropriate for graphene on SiO${}_2$.  We see that
this function does not approach zero as $\omega\to\infty$, and thus the $f$-sum diverges.  This divergence
is a direct consequence of the infinitely filled Fermi sea in the valence band.
We can in fact obtain an analytic expression for $\epsilon$ in the limit of large $\omega$; we find that
\begin{equation}
\epsilon(q,\omega)\approx 1+i\frac{\pi N\alpha q}{4\omega}.
\end{equation}
Note that this is not the expected form,
\begin{equation}
\epsilon(q,\omega)\approx 1-\frac{\omega_p^2}{\omega^2},
\end{equation}
where $\omega_p$ is the long-wavelength plasmon frequency\cite{HwangPRB2007}, $\omega_p\sim\sqrt{q}$.  We
see that, not only does the $f$-sum rule fail, but that the dielectric function has a nonstandard high-frequency
limit.  These seemingly strange results are due to the presence of an infinite Fermi sea, which would not
be present in an exact theory of graphene\cite{ThrockmortonArXiv}.  Because of this, interband scattering is
capable of scattering valence electrons from arbitrarily large negative energies into the conduction band.
To help illustrate this, we will recalculate the RPA dielectric function, keeping only the contributions
from intraband scattering.  This corresponds to keeping only the first terms in each of Eqs.\ (4) and (5)
of Ref.\ \onlinecite{HwangPRB2007}, so that
\begin{eqnarray}
\Pi^{+}_{IB}(q,\omega)&=&-\tfrac{1}{2}Ni\int\frac{d^2\vec{k}}{(2\pi)^2}\,\frac{f_{\vec{k},+}-f_{\vec{k}+\vec{q},+}}{\omega+v_F k-v_F|\vec{k}+\vec{q}|+i\eta} \cr
&\times&\left (1+\frac{\vec{k}\cdot(\vec{k}+\vec{q})}{k|\vec{k}+\vec{q}|}\right ), \\
\Pi^{-}_{IB}(q,\omega)&=&-\tfrac{1}{2}Ni\int\frac{d^2\vec{k}}{(2\pi)^2}\,\frac{f_{\vec{k},-}-f_{\vec{k}+\vec{q},-}}{\omega-v_F k+v_F|\vec{k}+\vec{q}|+i\eta} \cr
&\times&\left (1+\frac{\vec{k}\cdot(\vec{k}+\vec{q})}{k|\vec{k}+\vec{q}|}\right ),
\end{eqnarray}
where the $f_{\vec{k},\pm}$ are the Fermi occupation factors for electrons in the valence ($-$) and conduction ($+$)
bands.  If we now determine the resulting dielectric function, we find that the real and imaginary parts of the polarizability
for $|\omega|\leq v_Fq$ and $q\leq k_F$ may be written as
\begin{eqnarray}
\mbox{Re }\Pi_{IB}(q,\omega)&=&\frac{Nk_F}{2\pi v_F}xf_{IB}\left (\frac{q}{k_F},\frac{\omega}{v_F q}\right ), \\
\mbox{Im }\Pi_{IB}(q,\omega)&=&\frac{Nk_F}{2\pi v_F}xg_{IB}\left (\frac{q}{k_F},\frac{\omega}{v_F q}\right ),
\end{eqnarray}
where the functions $f(x,\nu)$ and $g(x,\nu)$ are given by
\begin{eqnarray}
&&f_{IB}(x,\nu)=\frac{1}{2\pi}\frac{1}{\sqrt{1-\nu^2}} \cr
&&\times\left (\int_{\nu<|t|\leq 1} dt\,\tanh^{-1}\left (\sqrt{\frac{1-\nu^2}{1-t^2}}\right )\sqrt{\left (\frac{2}{x}+t\right )-1}\right. \cr
&&+\left.\int_{-\nu}^\nu dt\,\tanh^{-1}\left (\sqrt{\frac{1-t^2}{1-\nu^2}}\right )\sqrt{\left (\frac{2}{x}+t\right )-1}\right ), \nonumber \\ \\
&&g_{IB}(x,\nu)=\frac{1}{8\sqrt{1-\nu^2}}\left [\left (\frac{2}{x}+\nu\right )\sqrt{\left (\frac{2}{x}+\nu\right )^2-1}\right. \cr
&&-\cosh^{-1}\left (\frac{2}{x}+\nu\right )-\left (\frac{2}{x}-\nu\right )\sqrt{\left (\frac{2}{x}-\nu\right )^2-1} \cr
&&+\left.\cosh^{-1}\left (\frac{2}{x}-\nu\right )\right ].
\end{eqnarray}
It turns out that the intraband contributions to the imaginary part are zero for $|\omega|>v_Fq$, so that we have in fact
completely specified it for all values of $\omega$.  We consider only the expressions for $q\leq k_F$ because we are interested
only in the low-wavelength behavior of the $f$-sum.

We now determine the long-wavelength (i.e., $q\ll k_F$) behavior of the $f$-sum.  To do this, we first determine the leading-order
behavior of the dielectric function.  The leading terms in $f$ and $g$ are
\begin{eqnarray}
f(x,\nu)&\approx&\frac{2}{\pi}\frac{1}{x\sqrt{1-\nu^2}}\left (\int_{0}^\nu dt\,\tanh^{-1}\left (\sqrt{\frac{1-t^2}{1-\nu^2}}\right )\right. \cr
&+&\left.\int_{\nu}^{1} dt\,\tanh^{-1}\left (\sqrt{\frac{1-\nu^2}{1-t^2}}\right )\right ), \\
g(x,\nu)&\approx&\frac{\nu}{x\sqrt{1-\nu^2}}.
\end{eqnarray}
We see that, at leading order, the dielectric function, as a function of $\frac{q}{k_F}$ and $\frac{\omega}{v_F q}$, goes as
$\frac{k_F}{q}$.

\begin{figure}[htb]
	\centering
		\includegraphics[width=0.49\columnwidth]{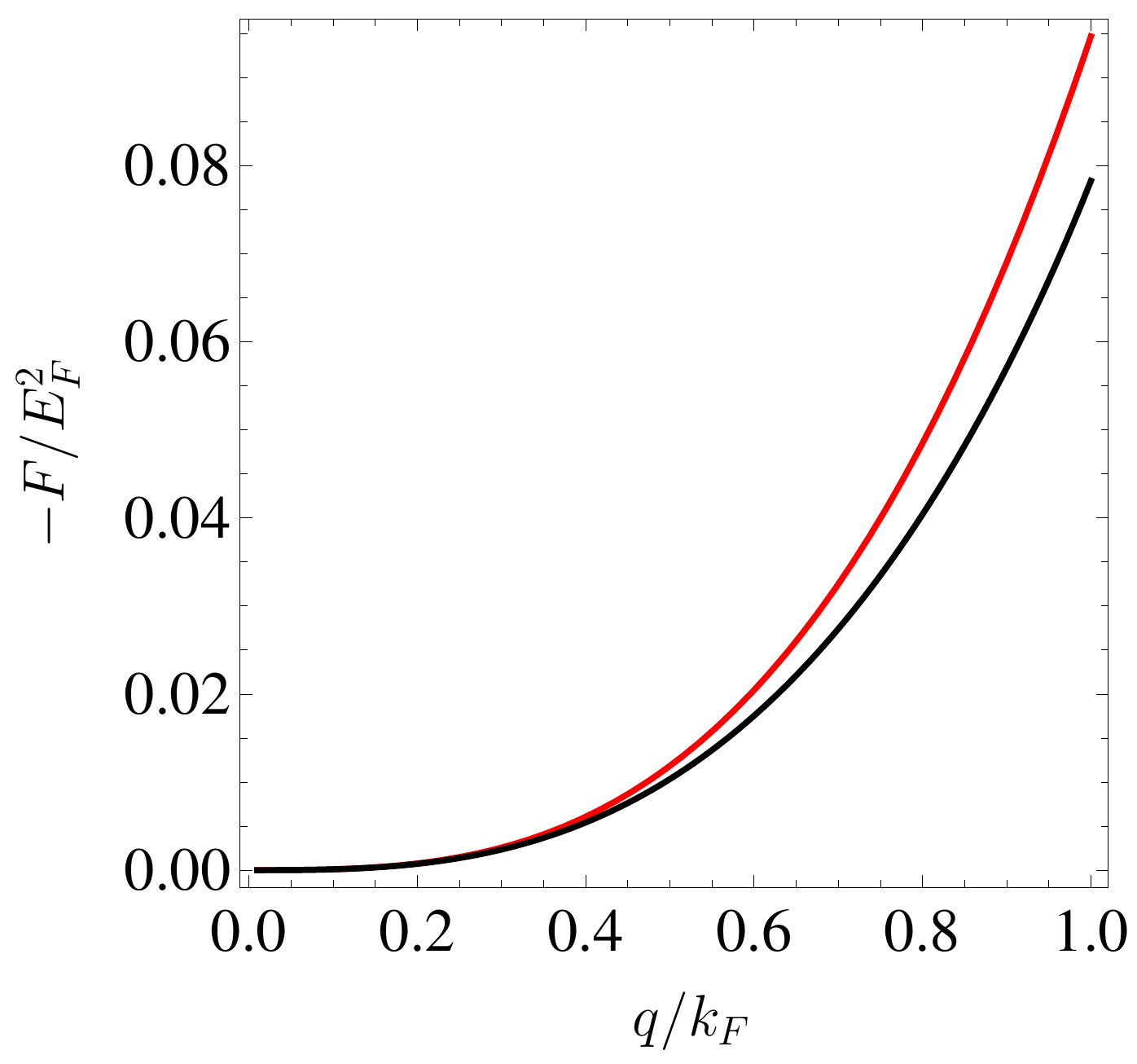}
		\includegraphics[width=0.49\columnwidth]{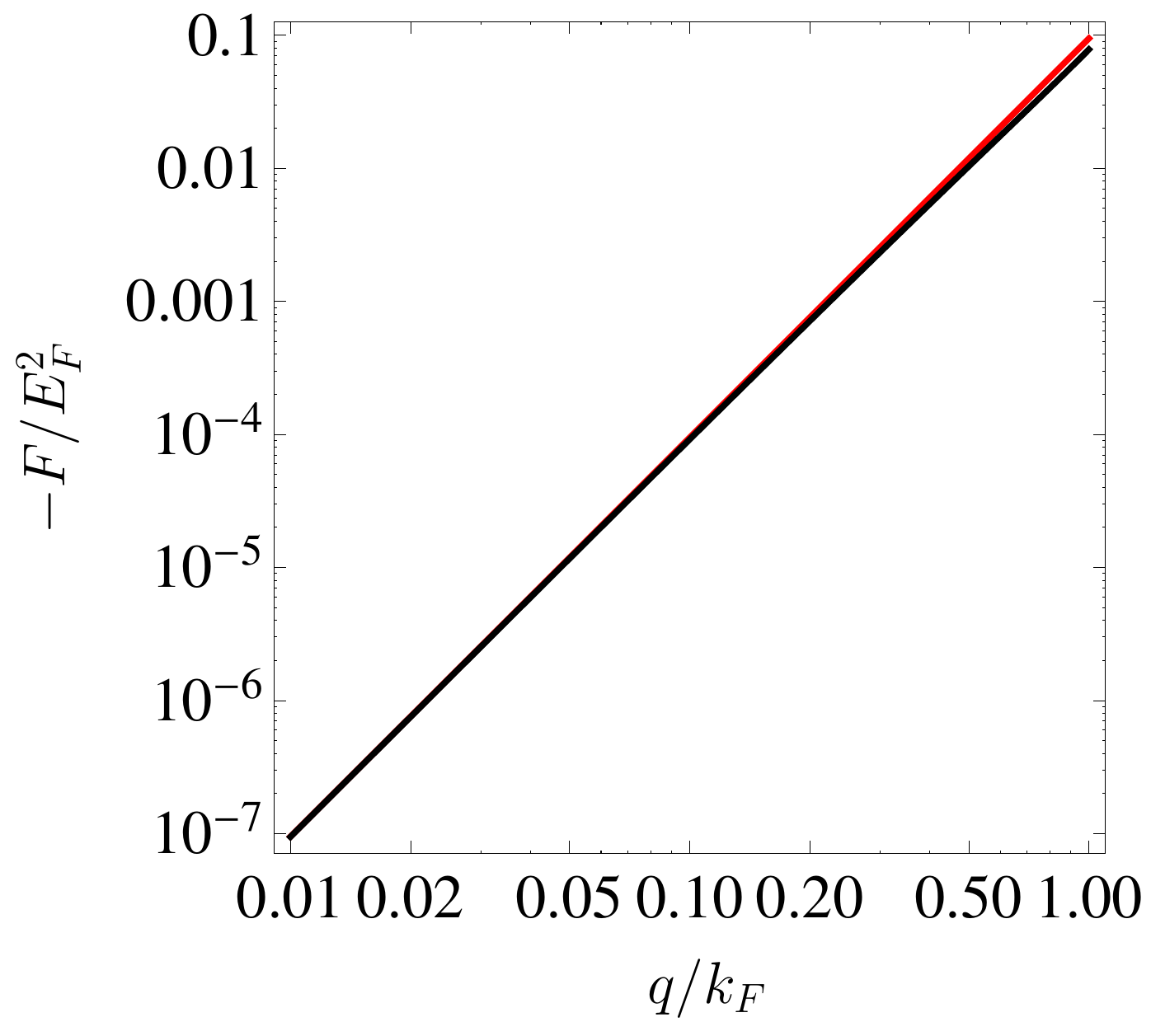}
	\caption{Plot of the negative of the $f$ sum, $F_{IB}$, computed from the dielectric function including only intraband scattering
	contributions, as a function of $q$ for $0\leq q\leq k_F$, both on linear	(left) and log-log (right) scales.  Here, we use $\alpha\approx 0.9$
	and $N=4$.}
	\label{Fig:fSumPlotsIB}
\end{figure}

Because we are able to write the dielectric function as a function only of $\frac{q}{k_F}$ and $\frac{\omega}{v_F q}$, and because
its imaginary part is nonzero only for $|\omega|\leq v_Fq$, we find that the $f$ sum can be written as
\begin{equation}
F_{IB}=-v_F^2 q^2\int_{0}^{1}d\nu\,\nu\,\frac{N\alpha g_{IB}(x,\nu)}{[1-N\alpha f_{IB}(x,\nu)]^2+[N\alpha g_{IB}(x,\nu)]^2}.
\end{equation}
If we now substitute the long-wavelength forms of $f$ and $g$ and perform a residual numerical integration, we find that the $f$ sum
goes as the cube of the wave vector; more precisely, it is given by
\begin{equation}
F_{IB}\approx -\frac{1}{3N\alpha}E_F^2\left (\frac{q}{k_F}\right )^3.
\end{equation}
The coefficient of $\tfrac{1}{3}$ is approximate; we obtained a value of $0.333333$.  We provide plots of both the exact and approximate
$f$ sum in Fig.\ \ref{Fig:fSumPlotsIB}; we deterimined the exact $f$ sum numerically.
We see that, if we only include the intraband scattering contribution to the dielectric function, then the $f$-sum becomes finite.  However,
we do not obtain the behavior expected from the $f$-sum rule stated earlier; if we did, then we should find that $F$ is linear in $q$.

We now calculate the $f$ sum with the full RPA dielectric function (i.e., we now also include interband scattering terms), but with an energy
cutoff $\Lambda$.  Because $\epsilon$ only depends on the dimensionless quantities, $x=\frac{q}{k_F}$ and $\nu=\frac{\omega}{E_F}$, the $f$ sum
may be rewritten as
\begin{equation}
F=E_F^2\int_{0}^{\Lambda/E_F}d\nu\,\nu\,\mbox{Im}\left [\frac{1}{\epsilon(x,\nu)}-1\right ].
\end{equation}

\begin{figure}[htb]
	\centering
		\includegraphics[width=\columnwidth]{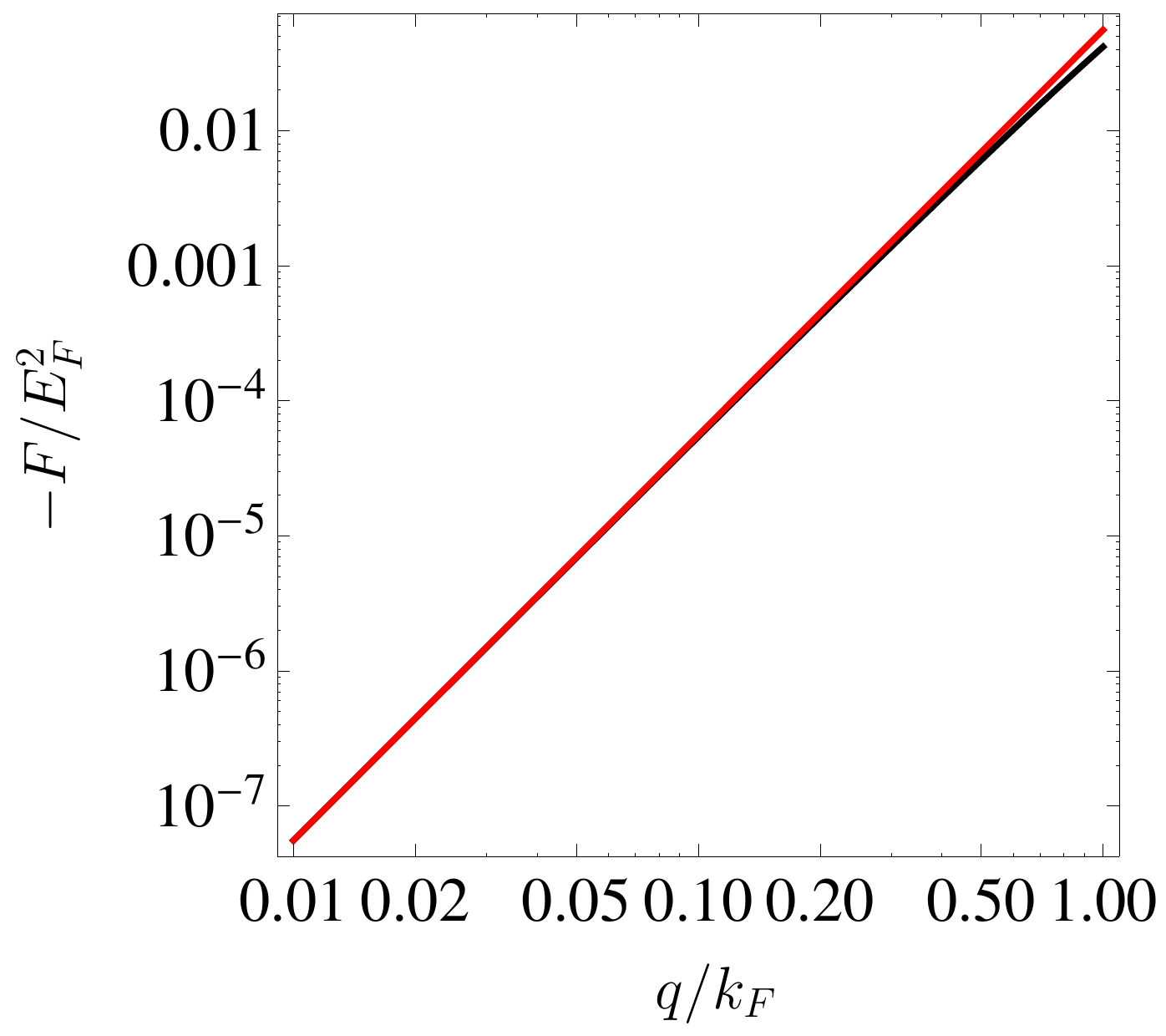}
	\caption{Plot of the exact intraband contribution to the $f$-sum (black) and the long-wavelength
approximation (red), Eq.\ \eqref{Eq:FSumIBFullRPA}, as a function of $q$ for $N=4$ and $\alpha\approx 0.9$.}
	\label{Fig:fSumIntrabandPart}
\end{figure}

While the $f$ sum for the full RPA dielectric function over all modes gives an infinite result, we will see that it
becomes finite if an energy cutoff is imposed on the integral. Let us first determine the contribution from frequencies
$0\leq\omega\leq v_F q$ (i.e., the ``intraband'' contribution as defined in Ref.\ \onlinecite{SabioPRB2008}).
While this integral must be found numerically in general, an analytic approximation exists for small $q$. In particular,
it can be shown that, at long wavelengths and with $\frac{\omega}{v_F q}$ held constant, the integrand of the $f$ sum
can be approximated as
\begin{equation}
\nu\,\mbox{Im}\left [\frac{1}{\epsilon(x,\nu)}-1\right ]\approx -\frac{\nu^2}{N\alpha x}\sqrt{x^2-\nu^2}.
\end{equation}
The resulting integral can be done analytically, and the result is
\begin{equation}
F=-\frac{\pi E_F^2}{16N\alpha}\left (\frac{q}{k_F}\right )^3. \label{Eq:FSumIBFullRPA}
\end{equation}
We plot this along with the exact result in Fig.\ \ref{Fig:fSumIntrabandPart}.

\begin{figure}[htb]
	\centering
		\includegraphics[width=\columnwidth]{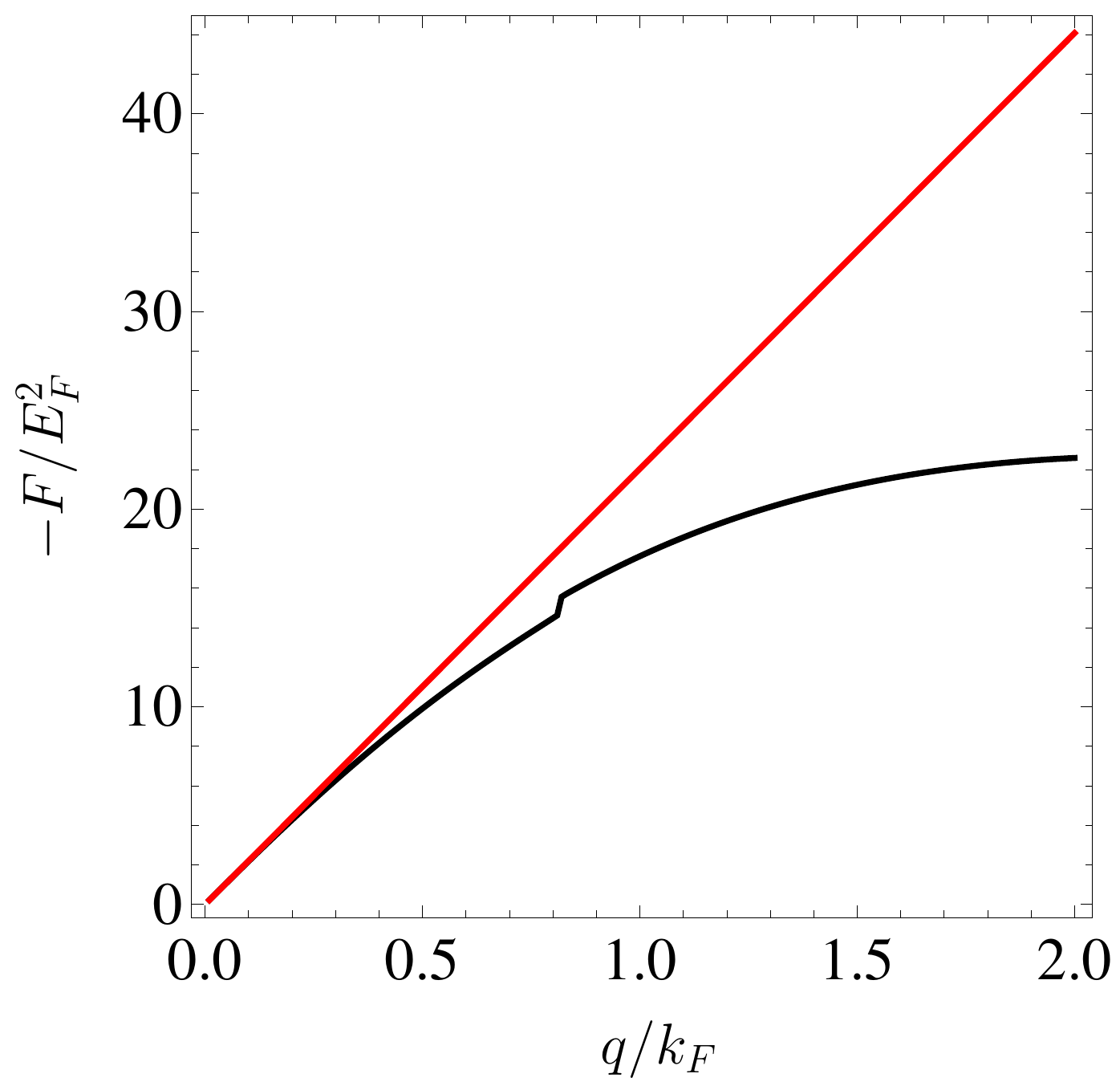}
	\caption{Plot of the $f$ sum (black) and its long-wavelength approximation (red) as a function
of $q$ for $\Lambda=10E_F$, $N=4$, and $\alpha\approx 0.9$.}
	\label{Fig:fSumPlotsLambda10EF}
\end{figure}

\begin{figure}[htb]
	\centering
		\includegraphics[width=\columnwidth]{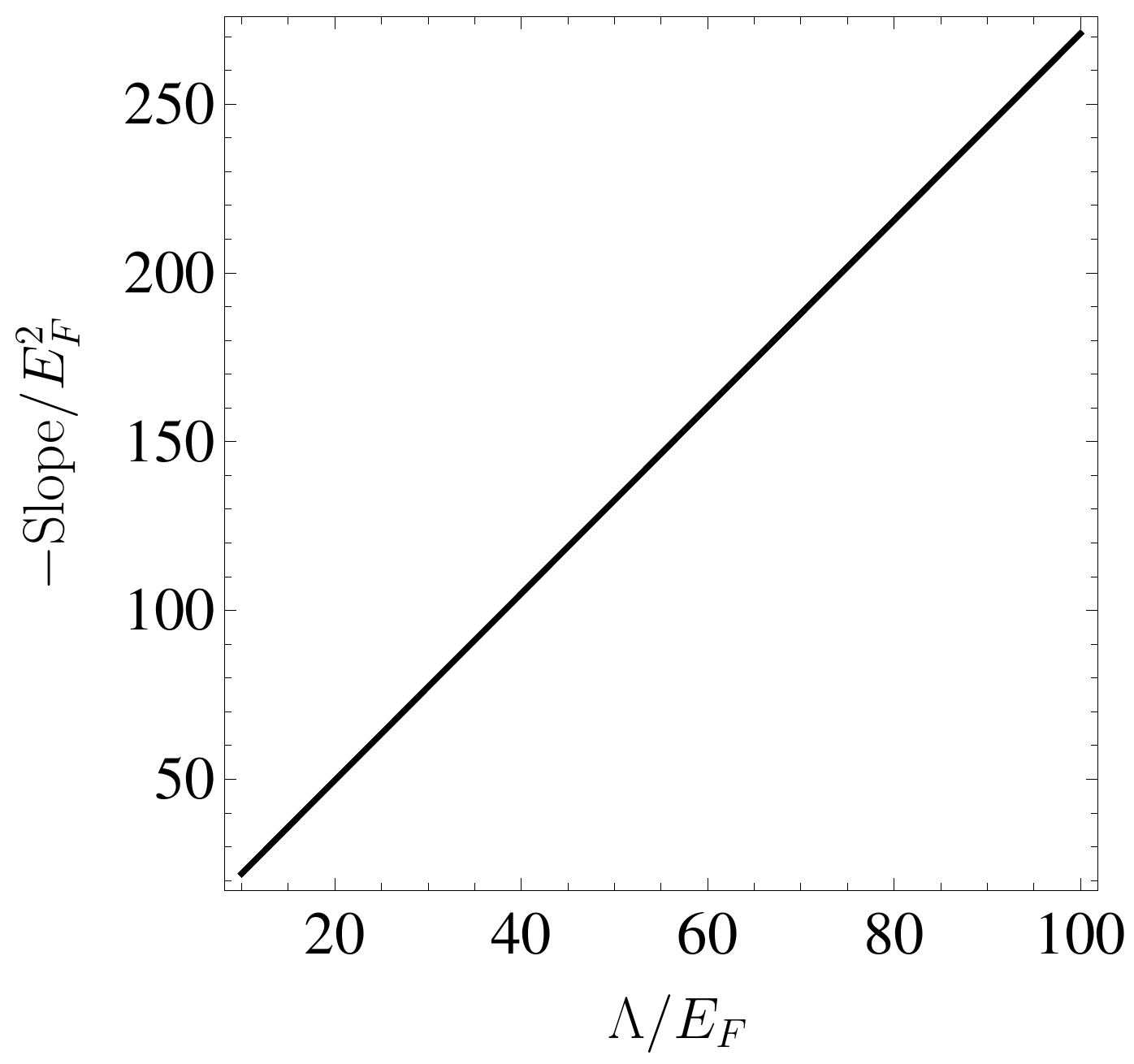}
	\caption{Plot of the slope of the $f$ sum at long wavelengths as a function of the energy
cutoff $\Lambda$ for $N=4$ and $\alpha\approx 0.9$.}
	\label{Fig:fSumSlopevsCutoff}
\end{figure}

We now consider the $f$ sum with an energy cutoff $\Lambda\gg E_F$.  We may write this $f$ sum as
\begin{equation}
F=E_F^2\int_{0}^{\Lambda/E_F}d\nu\,\nu\,\mbox{Im}\left [\frac{1}{\epsilon(x,\nu)}-1\right ].
\end{equation}
We thus see that the $f$ sum must be a function only of $\frac{\Lambda}{E_F}$ and $\frac{q}{k_F}$.
We provide a plot of the $f$ sum below for $\Lambda=10E_F$ in Fig.\ \ref{Fig:fSumPlotsLambda10EF}.
One can see that the $f$ sum appears to be linear in $q$ for small $q$.  We also investigate the
dependence of the slope of this approximate linear dependence as a function of $\Lambda$, and plot
the result in Fig.\ \ref{Fig:fSumSlopevsCutoff}.
The relationship between the cutoff and the slope appears to be linear.

\begin{figure}[htb]
	\centering
		\includegraphics[width=\columnwidth]{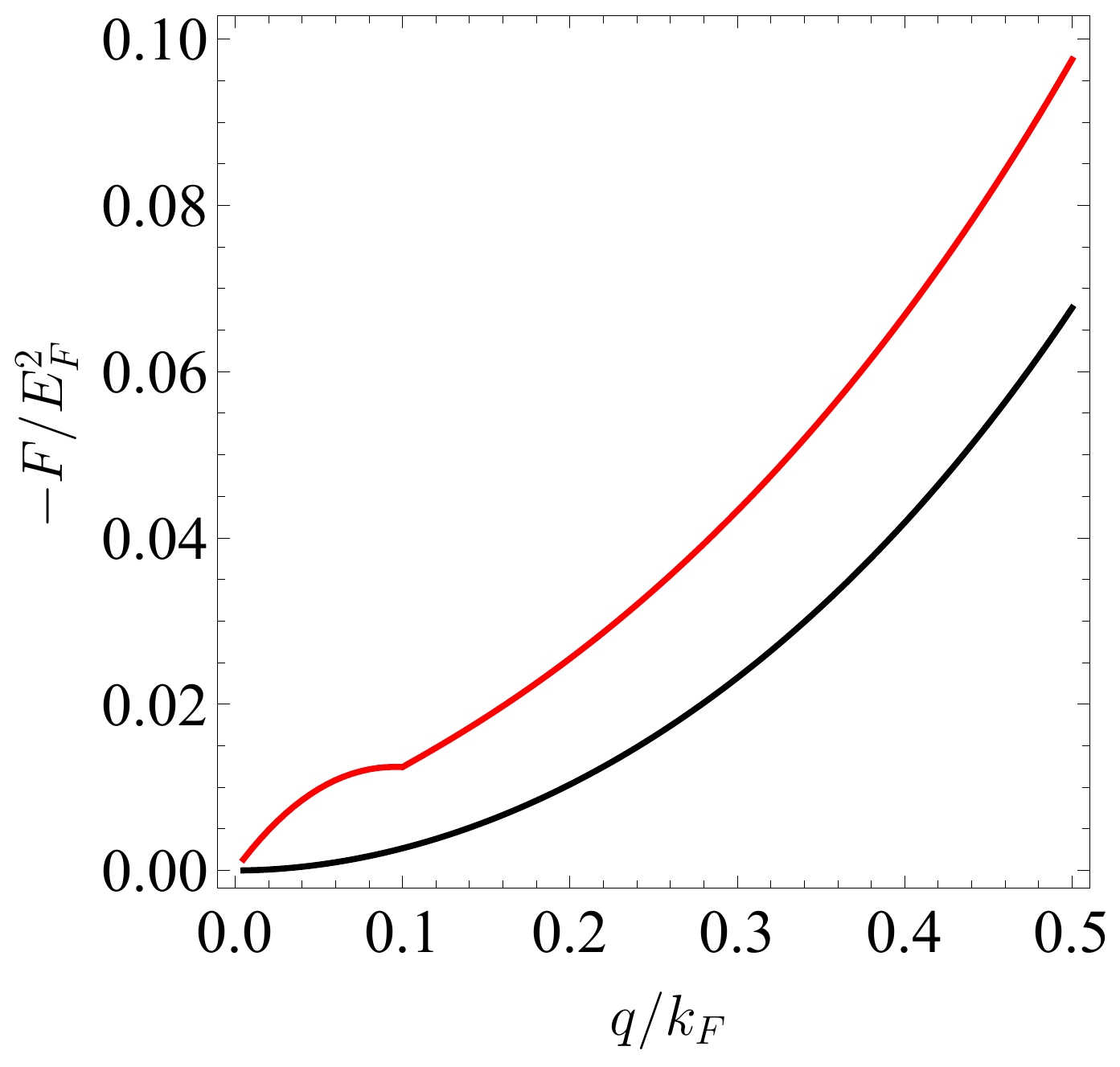}
	\caption{Plot of the $f$ sum at long wavelengths as a function of $q$ for $N=4$ and $\alpha\approx 0.9$,
and for $\Lambda=2E_F$ (black) and $\Lambda=2.1E_F$ (red).}
	\label{Fig:fSumCutoff20vs21}
\end{figure}

We also considered small cutoffs, equal or close to $2E_F$. This is the energy range within
which we find the plasmon modes when we determine them from the real part of the dielectric
function. If we do this with the energy cutoff $\Lambda=2E_F$, we find that the long-wavelength
behavior of the $f$-sum is quadratic in $q$.  However, if we increase the energy range even by
a very small amount, then we observe a linear behavior, again for small $q$.  We provide a plot
illustrating this effect in Fig.\ \ref{Fig:fSumCutoff20vs21}.

Everything that we have presented indicates that, for cutoffs $\Lambda>2E_F$, the $f$ sum is given
by $E_F^2$ times a linear function only of $\frac{\Lambda}{E_F}$ times $\frac{q}{k_F}$.  We thus
find that its long-wavelength behavior must be given by
\begin{equation}
F=-CE_F^2\left (\frac{\Lambda}{E_F}-2\right )\frac{q}{k_F},
\end{equation}
where $C\approx 2.20381$ in the case that $N=4$ and $\alpha\approx 0.9$.  This may be simplified to
\begin{equation}
F=-C(\Lambda-2E_F)v_F q.
\end{equation}
This form does produce a $\sqrt{q}$ dependence on $q$ for the long-wavelength plasmon frequency $\omega_p$.
Unfortunately, however, we find an effective plasmon frequency $\omega_p$ that is independent of the particle
density, contrary to the dependence found from the dielectric function directly\cite{HwangPRB2007}.  Note
that in developing the plasmon pole approximation for graphene in Sec.\ II of the main text, we completely
avoid the $f$-sum rule failure problem by demanding that the long-wavelength behavior of the effective
plasmon pole frequency $\omega_{\vec{q}}$ in Eq.\ \eqref{Eq:PPADielectric} go as $\omega_p$, where $\omega_p\sim\sqrt{q}$
is the actual long-wavelength graphene plasmon frequency.

\section{Hydrodynamic plasmon-pole approximation}
Since we have proposed the hydrodynamic plasmon-pole approximation as the most appropriate, as well as the
computationally simplest, model to be used in future graphene many-body calculations, we provide in this
appendix a critical comparison between the hydrodynamic approximation and the plasmon-pole approximation
for the frequency-dependent dielectric response.  We first define below the hydrodynamic [Eq.\ \eqref{Eq:DielectricFunc_Hyd}]
and plasmon-pole [Eq.\ \eqref{Eq:DielectricFunc_PPA}] dielectric functions:
\begin{eqnarray}
\epsilon_{\text{HD}}(\vec{q},\omega)&=&1-\frac{\omega_p^2}{\omega^2-v_F^2 q^2}=\frac{\omega^2-\omega_p^2-v_F^2 q^2}{\omega^2-v_F^2 q^2}, \nonumber \\ \label{Eq:DielectricFunc_Hyd_App} \\
\frac{1}{\epsilon_{\text{PPA}}(\vec{q},\omega)}&=&1+\frac{\omega_p^2}{\omega^2-\omega_{\vec{q}}^2}, \label{Eq:DielectricFunc_PPA_App}
\end{eqnarray}
or
\begin{equation}
\epsilon_{\text{PPA}}(\vec{q},\omega)=\frac{\omega^2-\omega_{\vec{q}}^2}{\omega^2-\omega_{\vec{q}}^2+\omega_p^2}. \label{Eq:DielectricFunc_PPA_App_Alt}
\end{equation}
Here, Eqs.\ \eqref{Eq:DielectricFunc_Hyd_App}--\eqref{Eq:DielectricFunc_PPA_App} correspond, respectively, to
Eqs.\ \eqref{Eq:DielectricFunc_Hyd}, Eqs.\ \eqref{Eq:PPADielectric} combined with \eqref{Eq:EffPlasmon_A}, and
Eq.\ \eqref{Eq:DielectricFunc_PPA}.

Now, the hydrodynamic PPA corresponds to [see Eq.\ \eqref{Eq:PPA_Hydrodynamic}] putting $\omega_{\vec{q}}^2=\omega_p^2+v_F^2 q^2$
into Eqs.\ \eqref{Eq:DielectricFunc_PPA_App} and \eqref{Eq:DielectricFunc_PPA_App_Alt}, leading to
\begin{equation}
\epsilon_{\text{PPA}}^{\text{HD}}(\vec{q},\omega)=\frac{\omega^2-\omega_p^2-v_F^2 q^2}{\omega^2-v_F^2 q^2}, \label{Eq:DielectricFunc_HydPPA_App}
\end{equation}
which is identical to the hydrodynamic dielectric function defined by Eq.\ \eqref{Eq:DielectricFunc_Hyd_App}.
Thus, the hydrodynamic dielectric function defined by Eq.\ \eqref{Eq:DielectricFunc_Hyd_App} exactly defines
the hydrodynamic approximation to the plasmon-pole approximation defined by Eq.\ \eqref{Eq:DielectricFunc_HydPPA_App}.

We emphasize that this identity between the hydrodynamic dielectric function and the plasmon-pole approximation
is achieved only after we make the hydrodynamic approximation to PPA [i.e., Eq.\ \eqref{Eq:PPA_Hydrodynamic}].  If
we compare the standard definitions of PPA [Eq.\ \eqref{Eq:DielectricFunc_PPA_App}] and hydrodynamics [Eq.\ \eqref{Eq:DielectricFunc_Hyd_App}],
\begin{equation}
\epsilon_{\text{PPA}}(\vec{q},\omega)=(\omega^2-\omega_{\vec{q}}^2)(\omega^2-\omega_{\vec{q}}^2+\omega_p^2)^{-1}
\end{equation}
and
\begin{equation}
\epsilon_{\text{HD}}(\vec{q},\omega)=(\omega^2-\omega_p^2-v_F^2 q^2)(\omega^2-v_F^2 q^2)^{-1},
\end{equation}
we see that $\epsilon(\vec{q},\tilde{\omega}_{\vec{q}})=0$ gives the following effective plasmon pole frequencies:
\begin{eqnarray}
\tilde{\omega}_{\vec{q}}^{\text{PPA}}&=&\omega_{\vec{q}}, \\
\tilde{\omega}_{\vec{q}}^{\text{HD}}&=&\sqrt{\omega_p^2+v_F^2 q^2}.
\end{eqnarray}
If we think of $\omega_{\vec{q}}$ as a plasma frequency, then the hydrodynamic approximation by virtue of having
the $v_F q$ term at second order guarantees that, for large $q$, the dispersion of $\tilde{\omega}_{\vec{q}}^{\text{HD}}$
goes as $v_F q$ following the graphene single-particle energy dispersion.  This is precisely the reason behind the
hydrodynamic approximation $\omega_{\vec{q}}=\tilde{\omega}_{\vec{q}}^{\text{HD}}=\sqrt{\omega_p^2+v_F^2 q^2}$
as in Eq.\ \eqref{Eq:PPA_Hydrodynamic} providing an excellent description for the effective plasmon pole frequency
$\omega_{\vec{q}}$: for small $q$, it produces the correct long-wavelength plasma frequency $\omega_p$ and, for
large $q$, it produces the correct graphene linear single-particle dispersion $v_F q$.

\end{document}